\documentclass[preprint,number,12pt]{elsarticle}
\journal{Information Processing Letters}

\newif\ifb\bfalse

\usepackage{verbatim}
\usepackage{url}
\usepackage{acro}
\usepackage{tikz}
\usepackage{subcaption} 
\usepackage{caption}
\usepackage{tkz-tab}
\usepackage{pgfbaseimage}
\usetikzlibrary{calc}
\usetikzlibrary{automata,positioning, snakes}

\newcommand{\third}{third}
\newcommand{\Third}{Third}

\date{\today{}}

\begin{document}

\begin{frontmatter}
	\title{%\sout{An attack against the Helios election}\\\sout{system that exploits re-voting}\\
    %\xout{Exploiting the Helios election system with re-voting}\\
    %\xout{A vulnerability in the Helios election system that exploits re-voting}\\
    %\xout{Exploiting $\{$a re-vote vulnerability, re-voting$\}$ in the Helios election system}\\
    %\xout{A $\{$re-voting exploit against, re-vote vulnerability in, re-vote exploit in/against$\}$ the Helios election system}\\
    %A re-vote exploit against the Helios election system\\[1em]
    Exploiting re-voting in the Helios election system
    }

	\author[vad]{Maxime Meyer%\corref{cor1}
}
%	\ead{research@maximemeyer.com}

	\author[lux]{Ben Smyth%\corref{cor2}
}
%	\ead{research@bensmyth.com}

	%\cortext[cor1]{Corresponding author}
	%\cortext[cor2]{Principal corresponding author}
	
  \address[vad]{Vade Secure Technology Inc., Montreal, Canada}
  
%	\address[hua]{Mathematical and Algorithmic Sciences Lab,\\
%	%France Research Center,
%	Huawei Technologies Co. Ltd., France}

\address[lux]{Interdisciplinary Centre for Security, Reliability and Trust,\\ University of Luxembourg, Luxembourg}

	\begin{abstract}
Election systems must ensure that representatives 
are chosen by voters.
Moreover, each voter should have equal influence. Traditionally, this has 
been achieved by permitting voters to cast at most one ballot. 
More recently, this has been achieved by tallying the last ballot cast by each voter. 
%We show that the Helios election system is unable to achieve this, 
We show this is not achieved by the Helios election system, 
because
an adversary can cause a ballot other than a voter's last 
to be tallied. Moreover, we show how the adversary can choose
the contents of such a ballot, thus the adversary can unduly influence
the selection of representatives.
\end{abstract}

%% We don't need keywords in arXiv version

%	\begin{keyword}
%	Security in digital systems \sep Analysis of algorithms \sep Cryptography \sep Electronic voting \sep Helios election system
	%\PACS 71.35.-y \sep 71.35.Lk \sep 71.36.+c I don't know what this is
%	\end{keyword}	
	
\end{frontmatter}

\thispagestyle{empty}

\section{Introduction}\label{intro}

An election is a decision-making procedure to choose representatives~\cite{Lijphart84,Saalfeld95,Gumbel05:StealThisVote,Alvarez10:ElectronicElections}.
Choices are made by voters, and this must be ensured by election systems,
as prescribed by 
the United Nations~\cite[Article 21]{UN:HumanRights},
the Organization for Security and Co‑operation in Europe~\cite[Paragraph 7.3]{OSCE:HumanRights}, and 
the Organization of American States~\cite[Article 23]{OAS:HumanRights}. 
These organisations also prescribe that systems must ensure that voters have equal 
influence in the decision~\cite{UN:HumanRights,OSCE:HumanRights,OAS:HumanRights}.
This has led to the emergence of the following 
eligibility and non-reusability
requirements~\cite{KR05:vote-privacy,Backes08:Voting}.\footnote{%
  Kremer \& Ryan capture both requirements in a single, informal definition, namely, 
  ``only...voters can vote, and only once"~\cite{KR05:vote-privacy}, whereas Backes 
  \emph{et al.} decouple that definition into ``only...voters can vote" and 
  ``every voter can vote only once"~\cite{Backes08:Voting}. (We refer to voters and 
  non-voters, whereas Kremer \& Ryan and Backes \emph{et al.} distinguish non-voters 
  from `legitimate voters' and `eligible voters.')}

\begin{itemize}
\item Eligibility. Choices are only made  by voters.
\end{itemize}

\noindent
Eligibility ensures that non-voters cannot (directly) influence 
the decision. For instance, national elections typically require
that voters are citizens of the nation, thus, eligibility forbids 
influence from foreign citizens.\footnote{We concede that non-voters 
may indirectly influence the decision, e.g., 
%non-voters may coerce voters to influence the decision.
voters may be swayed by disinformation~\cite{US:DisinformationAndDemocrary,UK:DisinformationAndDemocrary}.
}

\begin{comment}
Eligibility can be assured cryptographically (cf. eligibility verifiability~\cite{2014-election-verifiability}) 
or enforced by a \third\ party~\cite[\S2.2.3]{2014-election-verifiability}. In both cases, 
someone must be able to check that choices were made by voters, hence, choices must be 
authenticated.
\end{comment}

\begin{itemize}
\item Non-reusability. Only one choice of each voter has influence.
\end{itemize}

\noindent
Non-reusability ensures 
that each voter can contribute at most one choice, hence, voters
have equal influence. In addition, for verifiable elections~\cite{Benaloh85,Smyth10:ElectionVerifiability,Kusters11:UniversalVerifiability,2014-election-verifiability,Kiayias15},
non-reusability is useful to aid recovery from 
failure (since voters can 
``\emph{vote, verify, and revote until verification succeeds}"~\cite[\S1]{AdidaN06}).
\begin{comment}
Similarly to eligibility, non-reusability can be assured cryptographically 
%(cf. universal verifiability~\cite[\S4.2.2]{2014-election-verifiability}) 
or enforced by a \third\ party. %~\cite[\S2.2.2 \& \S2.2.3]{2014-election-verifiability}.
\end{comment}

Election systems have traditionally permitted each voter to cast at most one choice.
 More recent systems permit multiple choices (e.g.,~\cite{JCJ02,maaten2004towards,AdidaPereiraMarneffeQuisquater,Gjosteen2012,Cortier14:verifiability,Post10:re-voting})
and a voter's last choice should have influence.
We strengthen the aforementioned non-reusability requirement to capture such influence.

\begin{itemize}
\item Strong non-reusability. Only the last choice of each voter has influence.
\end{itemize}

\noindent
Strong non-reusability enables voters to change their choices, 
which provides flexibility, and aids education (since voters can 
``\emph{ask the help of anyone for submitting a random ballot, and then re-voting 
privately afterwards}"~\cite[\S3.3]{AdidaPereiraMarneffeQuisquater}). 
By comparison, (weak) non-reusability does not enable voters to change their choices, because that notion 
does not specify which ballot should have influence. Hence, it is permissible 
%\sout{for a voter's first vote to have influence, rather their last, for instance.} 
for a choice, other than the voter's last, to have influence. Consequently, voters cannot 
change their choices, because voters do not know which of their choices will have influence.
Thus, the notions of non-reusability by 
Kremer \& Ryan~\cite{KR05:vote-privacy} and Backes \emph{et al.}~\cite{Backes08:Voting} are 
too weak to analyse an interesting property of recent election systems; a slight strengthening
of their notions is necessary.

Eligibility and non-reusability are fundamental requirements of election systems~\cite{UN:HumanRights,OSCE:HumanRights,OAS:HumanRights}, as-is
strong non-reusability when voters are permitted to change their choices. These requirements
all assume that the adversary's capabilities are limited to controlling the communication 
channel and that the election system is operated in the prescribed manner, hence, they 
are not intended to exclude attacks that arise when the election system is subverted by
the adversary (to authenticate non-voter ballots, for instance). By comparison, verifiability
requirements assume the election system has been subverted and are intended
to enable the detection, rather than exclusion, of attacks~\cite{Benaloh85,Smyth10:ElectionVerifiability,Kusters11:UniversalVerifiability,2014-election-verifiability,Kiayias15}. 
It follows that an election system
that satisfies eligibility, non-reusability, and strong non-reusability is invulnerable to
attacks against those requirements when the election system is operated in the prescribed 
manner, whereas a verifiable system might be vulnerable to attacks, but those attacks can 
be detected. Thus, eligibility, non-reusability, and strong non-reusability should be satisfied
regardless of whether verifiability is, because election systems operated in the prescribed
manner should prevent attacks by network adversaries, 
rather than just enabling attack detection. 

%Hence, ballots must be authenticated. Some voting systems rely on a \third\ party to
%authenticate voters' ballots, whereas others rely on a \third\ party to issue credentials 
%to voters and use cryptography to ensure that only voters can construct authorised ballots. 
%In the absence of a \third\ party, auditing can be used to check whether only voters' ballots
%were tallied, but that goes beyond the scope of eligibility, 

%Ultimately, auditing is required to ensure that only voters' ballots are authenticated or that
%only voters are issued with credentials. Such auditing is beyond the scope of our eligibility
%requirement, which assumes ballots are correctly authenticated. 

We analyse Helios~\cite{AdidaPereiraMarneffeQuisquater}: an open-source, web-based election system,%
\footnote{\url{https://vote.heliosvoting.org}, accessed 11 Aug 2017.}\
which has been used %in the real-world~\cite{IACRHelios10b,IACRHelios10,AdidaPereiraMarneffeQuisquater,Adida09:HeliosPrinceton}.
by the International Association of Cryptologic Research (IACR),
the ACM, the Catholic University of Louvain, 
and Princeton University~\cite{Olivier16:Helios}.
Helios uses %OAuth~\cite{Citation-Needed} 
a \third\ party 
%for authentication and
%\footnote{\url{https://github.com/benadida/helios-server/tree/master/helios_auth/auth_systems}, accessed 4 Aug 2015.} 
\begin{comment}
we show that %the use of OAuth by Helios 
the authentication mechanism is 
\end{comment}
to authenticate voters' ballots, which suffices for eligibility,
assuming the \third\ party is trusted. Authenticated ballots are
listed alongside voter identities and at most one ballot is listed
alongside each identity, which suffices for non-reusability. 
Any other ballots are archived. Auditing can be used to statistically 
determine whether non-voter ballots are incorrectly authenticated by an 
untrusted third party or whether unauthenticated ballots are listed. %(without determining which). 
For instance, voters can be asked whether the ballot 
alongside their identity is theirs, to determine if the ballot
was incorrectly authenticated or simply unauthenticated. %(without determining which). 
Albeit cooperation and honesty of some voters is required for auditing, 
and sufficiently many malicious voters can 
%cause audits to incorrectly fail. 
manipulate audits.

\paragraph{Contribution}
We show that %the use of OAuth by Helios 
the archiving mechanism used by Helios is 
insufficient to ensure strong non-reusability,
in the presence of an adversary that is able to delay 
messages sent on the network.
%, because %an adversary 
In particular, the adversary 
can cause a choice other than a voter's last 
to be counted. Moreover, we show how the adversary can 
pick the choice, in a poll station with a 
malicious election supervisor. Although 
malice can be detected by
voters that perform verifiability checks once voting closes, 
recovery is only possible before tallying commences.

\section{Analysis of Helios}

\subsection{Protocol description}

\begin{figure}
    \centering
	\begin{subfigure}[b]{0.95\textwidth}
        \centering
        \resizebox{\linewidth}{!}{
	    		\begin{tikzpicture}
				\draw (14, 10) rectangle +(-3, -4);
				\node[anchor=north west] at (11, 10) {Bulletin Board};

				\node (voter 1) at (0, 8) {\includegraphics[width=50pt]{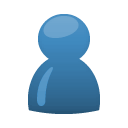}};
  				\node (vote 1) at (2, 8) {\includegraphics[width=40pt]{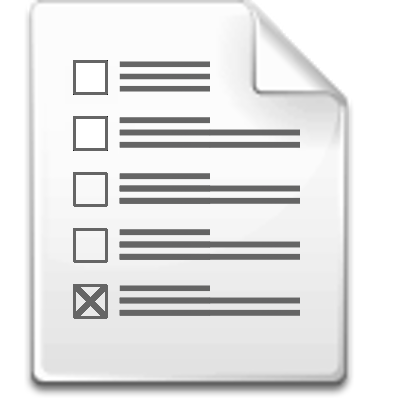}};
				\node at ($(voter 1) + (0, -1.5)$) {Voter};  			
  			
  				\node (voter 1) at (0, 8) {\includegraphics[width=50pt]{voterblue.png}};
				\node (ballot 1) at (6, 8) {\includegraphics[width=40pt]{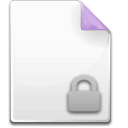}};
				\draw[->] (vote 1) -- node[above] {1} (ballot 1);
				%\node at ($(ballot 1) + (-0.2, 0.1)$) {\textcolor{red}{$1$}};
				
				\node (ballot 0) at (12, 8) {\includegraphics[width=40pt]{lockedfilepurple.png}};
				%\node at ($(ballot 0) + (-0.2, 0.1)$) {\textcolor{red}{$1$}};
				\draw[->] (ballot 1) -- node[above] {2} (ballot 0);		
			\end{tikzpicture}
		}
		\caption{Casting a ballot}
    		\label{fig:flowa}
	\end{subfigure}

	\par\bigskip	
	
	\begin{subfigure}[b]{0.95\textwidth}
        \centering
        \resizebox{\linewidth}{!}{
	    		\begin{tikzpicture}
				\draw (14, 10) rectangle +(-3, -4);
				\node[anchor=north west] at (11, 10) {Bulletin Board};		 				 			
  				\node (voter 1) at (7, 10) {\includegraphics[width=50pt]{voterblue.png}};
				\node (server) at (0, 7) {\includegraphics[width=50pt]{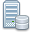}};
				\node at ($(voter 1) + (0, -1.5)$) {Voter};
				\node at ($(server) + (0, -1.5)$) {\Third\ Party};
				
				\node (c0) at (11.5, 8.5) {\includegraphics[width=30pt]{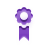}};
 				%\node at ($(c0) + (0, 0.1)$) {\textcolor{black}{$0$}};
 				\node (ballot 0) at (12.7, 7.2) {\includegraphics[width=40pt]{lockedfilepurple.png}};
				%\node at ($(ballot 0) + (-0.2, 0.1)$) {\textcolor{red}{$1$}};
				
				\draw[->] (voter 1) -- node[above, sloped] {5} (c0); 				
 				
				\draw[->] ($(voter 1) + (-1.3, -0.1)$) -- node[above, sloped] {3} ($(server) + (1.3,0.2)$);
				\draw ($(server) + (1.3,-0.2)$) -- node[below, sloped] (c1) {} ($(voter 1) + (-1.3, -0.5)$);
				\node at ($(c1) + (0.8, -0.1)$) {\includegraphics[width=30pt]{sealpurple.png}};
				\draw[->] ($(server) + (1.3,-0.2)$) -- node[below, sloped] (c1) {$4$} ($(voter 1) + (-1.3, -0.5)$);
				\draw (14, 10) rectangle +(-3, -4);
				
				\draw[<-] ($(server) + (1.3,-0.4)$) -- node[above] {6} (10.9, 6.6); 
				\draw[->] ($(server) + (1.3,-0.9)$) -- node[below] {7} (10.9, 6.1); 
				
				%\node at ($(c1) + (0, 0.1)$) {\textcolor{black}{$0$}};
			\end{tikzpicture}
		}
		\caption{Authenticating the ballot}
    		\label{fig:flowb}
	\end{subfigure}

	\par\bigskip
	
	\begin{subfigure}[b]{0.95\textwidth}
        \centering
        \resizebox{\linewidth}{!}{
	    		\begin{tikzpicture}	
			  \draw (12, 11) rectangle +(-6, -8);

			  \node[anchor=north west] at (6, 11) {Bulletin Board};
			  
			  \node (admins) at (0, 8) {\includegraphics[width=50pt]{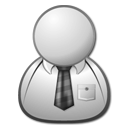}};
			  \node at ($(admins) + (0, -1.5)$) {Administrator}; 
			  \node (c3) at (9, 7) {\includegraphics[width=30pt]{sealpurple.png}};
			  \node (bb 1) at (8, 8) {\includegraphics[width=40pt]{lockedfilepurple.png}}; 
			  
			  %\node at ($(bb 1) + (-0.2, 0.1)$) {\textcolor{red}{$1$}};
			  %\node at ($(c3) + (0, 0.1)$) {\textcolor{black}{$0$}};
			
			  \node (ballots) at (10, 10) {\includegraphics[width=40pt]{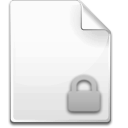}}; 
			  \node at ($(ballots) + (0.25, -0.25)$) {\includegraphics[width=40pt]{lockedfile.png}}; 
			  \node at ($(ballots) + (0.5, -0.5)$) {\includegraphics[width=40pt]{lockedfile.png}};
			
			  \node (sum ballot) at (8, 4) {\includegraphics[width=40pt]{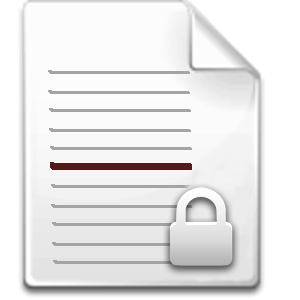}};
			  \node (sum) at ($ (sum ballot) + (2.5, 0) $) {$\bigoplus$};
			  %\node at ($ (sum ballot) + (-0.2, 0.1) $) {$\Sigma$};
			  \draw (bb 1) -- (bb 1 -| sum) -- (ballots -| sum);
			  \draw[->] (ballots) -- (ballots -| sum) -- node [right] {8}  (sum);
			  \draw[->] (sum) -- (sum ballot);
			  
			  \node at ($(ballots) + (0.5, -0.5)$) {\includegraphics[width=40pt]{lockedfile.png}};
			  
			  \node (res) at (1, 4) {\includegraphics[width=40pt]{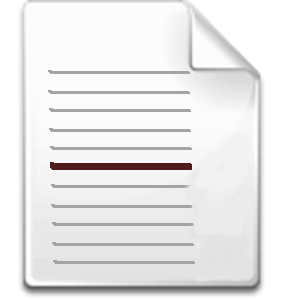}};
			  \draw[->] (sum ballot) -- node [below] {10} (res);
			  \draw[->] (admins) -- node [above] {9} (sum ballot);
			\end{tikzpicture}
		}
		\caption{Tallying}
    		\label{fig:flowc}
		\end{subfigure}

	\caption{Helios protocol flow} \label{fig:flow}
\end{figure}

An execution of Helios (Figure \ref{fig:flow}) proceeds as follows.
First, a voter casts a ballot for their choice: 
the voter encrypts their choice (1) and 
sends their encrypted choice to the bulletin board (2). %(Choice $1$ is shown in the figure.)
Secondly, the voter authenticates their encrypted choice to the bulletin board, 
to prove they are indeed a voter.
The authentication process uses OAuth~\cite{RFC6749},\footnote{%
  Other authentication methods are also supported.} 
which is reliant on a \third\ party.
The process proceeds as follows.
The voter authenticates to a \third\ party (3),
the \third\ party generates a token for the voter (4),
the voter sends the token to the bulletin board (5),
and the bulletin board relays the token to the \third\ party (6).
The \third\ party checks whether the token is valid and 
 notifies the bulletin board of the token's validity (7).
If the token is valid, then the bulletin board accepts the voter's encrypted choice.
Hence, the bulletin board contains the voter's authenticated encrypted choice.
In addition, the bulletin board archives any encrypted choice previously accepted for that voter,\footnote{See \texttt{Voter.last\_cast\_vote()} in \url{https://github.com/benadida/helios-server/blob/9fa74a2bef41c0c344f1c9a6f1c28a36f93347ea/helios/models.py}, accessed 11 Aug 2017.}
%https://github.com/benadida/helios-server/blob/master/helios/models.py
which is intended to ensure that only the last choice of each voter has influence.
Finally, the bulletin board homomorphically combines the accepted encrypted choices (8), 
the administrator decrypts the homomorphic combination (9), and the bulletin board 
reveals those decrypted choices (10).
Helios satisfies eligibility, because encrypted choices are only accepted by the bulletin board
%if the voter presents a token demonstrating that they authenticated to the \third\ party
when accompanied by a token authenticating the voter that constructed the encrypted choice. 
Moreover, Helios satisfies non-reusability too, because,
upon acceptance, the bulletin board archives any encrypted 
choice previously accepted for that voter. 
But, this is insufficient for strong non-reusability.

\subsection{Vulnerability}

\begin{figure}
	\centering
	
	\begin{subfigure}[b]{\textwidth}
        \centering
        \resizebox{\linewidth}{!}{
	    		\begin{tikzpicture}
				\draw (14, 9) rectangle +(-3, -4);
				\node[anchor=north west] at (11, 9) {Bulletin Board};		 				 			
  				\node (voter 1) at (5, 7) {\includegraphics[width=50pt]{voterblue.png}}; 
				\node (server) at (-2, 7) {\includegraphics[width=50pt]{server.png}};
				\node at ($(voter 1) + (0, -1.5)$) {Voter};
				\node at ($(server) + (0, -1.5)$) {\Third\ Party};
				
				\node (adversary) at (9, 4) {\includegraphics[width=50pt]{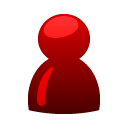}};
				\node at ($(adversary) + (0, -1.5)$) {Adversary};
				
				%\node (c1) at (2, 6) {\pgfuseimage{certificate}};;
				\node (ballot 0) at (12, 7) {\includegraphics[width=40pt]{lockedfilepurple.png}};
				%\node at ($(ballot 0) + (-0.2, 0.1)$) {\textcolor{red}{$0$}};
												
				\draw[->] ($(voter 1) + (-1.3, 0.2)$) -- node[above] {1} ($(server) + (1.3,0.2)$);
				\draw[->] ($(server) + (1.3,-0.2)$) -- node[below] (c1) {\includegraphics[width=30pt]{sealpurple.png}} ($(voter 1) + (-1.3, -0.2)$);
				\node at ($(c1) + (-1,0)$) {2\ ,};
				\draw[->] ($(voter 1) + (1.3,0)$) -- node[above] (c2) {\includegraphics[width=30pt]{sealpurple.png}} (9,7);
				\node at ($(c2) + (-1,0)$) {3\ ,};
				\draw[red, <-, decorate,
					decoration={
						snake,
						amplitude=.4mm,
						segment length=2mm,
						pre length=2mm
						}] (9, 6.8) -- node[right] {4}  (adversary);	
			\end{tikzpicture}
		}
		\caption{Token interception}
    		\label{fig:attacka}
	\end{subfigure}

	\par\bigskip
	
	\begin{subfigure}[b]{\textwidth}
        \centering
        \resizebox{\linewidth}{!}{
	    		\begin{tikzpicture}
				\draw (16, 10) rectangle +(-5, -4);
				\node[anchor=north west] at (11, 10) {Bulletin Board};		 				 			
  				\node (voter 1) at (7, 7) {\includegraphics[width=50pt]{voterblue.png}};
				\node (server) at (0, 7) {\includegraphics[width=50pt]{server.png}};
				\node at ($(voter 1) + (0, -1.5)$) {Voter};
				\node at ($(server) + (0, -1.5)$) {\Third\ Party};
				
				\node (c0) at (13, 7) {\includegraphics[width=30pt]{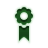}};
				\node (ballot 0) at (12, 8) {\includegraphics[width=40pt]{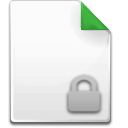}};
				%\node at ($(ballot 0) + (-0.2, 0.1)$) {\textcolor{red}{$1$}};
				\node (ballot 1) at (15, 9) {\includegraphics[width=40pt]{lockedfilepurple.png}};
				%\node at ($(ballot 1) + (-0.2, 0.1)$) {\textcolor{red}{$0$}};				
				\draw[->] ($(voter 1) + (1, 1)$) -- node[above] {5} (ballot 0); 									
				\draw[->] ($(voter 1) + (-1.3, 0.2)$) -- node[above] {6}  ($(server) + (1.3,0.2)$);
				\draw[->] ($(server) + (1.3,-0.2)$) -- node[below] (c1) {\includegraphics[width=30pt]{sealgreen.png}} ($(voter 1) + (-1.3, -0.2)$);
				\node at ($(c1) + (-1,0)$) {7\ ,};
				
				\draw[->] (voter 1) -- node[above] {8} (c0); 
			\end{tikzpicture}
		}
		\caption{Casting a second ballot}
    		\label{fig:attackb}
	\end{subfigure}

	\par\bigskip
	
	\begin{subfigure}[b]{\textwidth}
        \centering
        \resizebox{\linewidth}{!}{
	    		\begin{tikzpicture}
				\draw (11, 10) rectangle +(-7, -4);
				\node[anchor=north west] at (4, 10) {Bulletin Board};		 
				
				\node (server) at (-3, 7) {\includegraphics[width=50pt]{voterblue.png}};
				\node at ($(server) + (0, -1.5)$) {Voter};				 
  				\node (adversary) at (0, 4) {\includegraphics[width=50pt]{voterred.png}};
				\node at ($(adversary) + (0, -1.5)$) {Adversary};
				
				\node (c0) at (6, 7) {\includegraphics[width=30pt]{sealpurple.png}};
 				\node (c1) at (9, 8) {\includegraphics[width=30pt]{sealgreen.png}};
 				\node (ballot 0) at (5, 8) {\includegraphics[width=40pt]{lockedfilepurple.png}};
				%\node at ($(ballot 0) + (-0.2, 0.1)$) {\textcolor{red}{$0$}};
				\node (ballot 1) at (8, 9) {\includegraphics[width=40pt]{lockedfilegreen.png}};
				%\node at ($(ballot 1) + (-0.2, 0.1)$) {\textcolor{red}{$1$}};				
				\node at ($(c1) + (1.1,-0.5)$) {10};
				\draw[red] ($(c1) + (1,-1)$) -- ($(ballot 1) + (-0.9,0.9)$);	
				
				\draw[dashed] ($(server) + (1.3,0)$) -- node[above] {\includegraphics[width=30pt]{sealpurple.png}} (0,7);
				\draw[->] (0,7) -- node[above] {9} (c0);
				\draw[black!40!green, <-, decorate,
					decoration={
						snake,
						amplitude=.4mm,
						segment length=2mm,
						pre length=2mm
						}] (0, 6.8) -- (adversary);			
			\end{tikzpicture}
		}
		\caption{Release intercepted token}
    		\label{fig:attackc}
	\end{subfigure}

	\caption{Helios exploit flow} \label{fig:attack}
\end{figure}

The flow of our exploit initially corresponds to an honest execution: 
a voter casts a ballot for their choice, as per Figure~\ref{fig:flowa}. 
The remaining steps (Figure~\ref{fig:attack}) proceed as follows.
First, the adversary intercepts a voter's token:
the voter authenticates to a \third\ party (1), 
receives an authentication token (2), and sends the token to the bulletin board (3), 
but it is intercepted by the adversary (4).\footnote{%
An adversary can intercept packets even when they are encrypted.
For example, packets sent over a TLS connection, i.e., encrypted packets, can be intercepted.
Intercepting a TLS packet prevents further data from being received on \emph{that} TLS connection (until the packet is released), but data may be received on other TLS connections (of which there are many), because TLS does not guarantee ordering of messages between connections. (Multiple TLS connections are maintained to reduce latency.) Hence, TLS does not prevent further communication between the voter and the bulletin board.
}
(We indicate the ballot-token relation by colouring the top right-hand corner of the ballot and the token in purple.)
Thus, the bulletin board contains an unauthenticated encrypted choice 
and is awaiting an authentication token for that encrypted choice. 
Next, the adversary waits until the voter casts another encrypted choice (5), authenticates with the \third\ party (6), receives a token (7), and sends the token to the bulletin board (8).
(We indicate the ballot-token relation using green colouring.)
Thus, the bulletin board can authenticate the voter's second ballot.
Finally, the adversary releases the intercepted token and it is received by the bulletin board (9).
Thus, the bulletin board will accept the voter's first ballot, and archive the voter's second ballot (10). 
Consequently, the voter's first choice is counted, 
rather than their second. Hence, strong non-reusability is not satisfied, because only the last choice of the voter should have influence, which is not the case.

\paragraph{Video demonstration} The exploit is demonstrated 
in a supporting video \cite{Smyth14:HeliosEligibilityVideo}.\\

Helios developers Ben Adida and Olivier Pereira acknowledge the existence of this vulnerability, but they contend it would be detected.\footnote{Email communication, April 2014.} 
We will discuss detection mechanisms in the following section.

\subsection{Impact}

Let us now consider %how an adversary might unduly influence
the possibility of an adversary unduly influencing
an election's 
outcome, in settings where Helios is deployed in voting terminals located at poll stations.
In such settings, a malicious election supervisor could offer to demonstrate the Helios 
system to a voter, under the guise of education. During the demonstration, the supervisor 
could suggest that the voter selects a particular choice. This should not cause suspicion, 
because Helios is intended to permit voters to change their choices (\S\ref{intro}). Once the voter casts 
the demonstration ballot, it could be intercepted, perhaps by a router in the polling station 
that the supervisor controls. After the demonstration, the supervisor could instruct the 
voter to re-vote in private. 
Once the voter leaves the poll station, the intercepted ballot could be released. 
Consequently, the supervisor's choice is tallied, rather than the voter's, thereby
demonstrating the possibility of undue influence.

We acknowledge that the voter can discover that malice has taken place, because Helios
satisfies individual verifiability~\cite{2014-election-verifiability,Smyth10:ElectionVerifiability} 
and the voter can check whether the bulletin board accepted their second encrypted choice.
However, it is well-known that many voters do not perform checks necessary for verifiability
and voting systems rely on checks being performed by sufficiently many diligent 
voters~\cite[\S2.1.6]{Ryan17:sok-voting-challenges}. Thus, the exploit is particularly 
effective against voters that do not perform checks.
Moreover, even if malice is detected, recovery is only possible 
when a voter successfully casts another encrypted choice (before tallying), 
hence, the exploit can also be effective against voters that detect malice.
%The adversary can improve effectiveness 
Effectiveness can be improved 
by releasing the intercepted ballot just before voting 
closes, this not only reduces the voter's opportunity to detect malice, but also 
forces the voter to convince officials that they should be able to cast another 
ballot after voting closes, which is problematic, because there is no convincing 
evidence that any malpractice has taken place.
Once 
tallying commences, the voter cannot recover, furthermore,
\begin{comment}
there is no convincing evidence that any 
malpractice has taken place% (cf. accountability~\cite{Kusters10})%
%. Indeed, malice cannot be attributed to a particular party (cf. accountability~\cite{Kusters10})
%. Indeed, the \third\ party and the bulletin board both behaved correctly, and the evidence 
%  seemingly suggests that the voter is cheating. 
, thus, victims have little recourse.
\end{comment}
given the absence of any evidence,  
victims have little recourse.

We believe that verification checks should serve as a last line of defence and 
that election systems should prevent many attacks, rather than merely being able to 
detect them (especially as detection does not imply the ability to recover).
%Voters should not be satisfied by voting systems that necessitate checks by all voters 
%(or even a majority of voters), since this is unrealistic.
Hence, 
we believe eligibility and non-reusability are
worthy of study independently of verifiability.

\subsection{Fixes}

We can patch the vulnerability
 by checking authentication token timestamps, timestamping ballots, coupling encrypted choices with counters, or proving knowledge of earlier encrypted choices \emph{\`a la} Clarkson, Chong \& Myers~\cite[\S3.3]{CCM08}. 
We favour solutions using timestamps, since the other approaches require the voter to maintain state. 
Moreover, timestamps have been acknowledged as a possibility for 
a fix by Helios developer Olivier Pereira.\footnote{Email communication, April 2014.}
We concede that timestamps increase the attack surface, since an adversary 
may tamper with clocks. But, we stress that the \third\ party is already 
assumed to construct authentication tokens correctly and that voters are 
already assumed to construct ballots correctly (or, at least, audit ballots to 
increase confidence of correct construction),\footnote{%
  Auditing ballots provides assurance that ballots (constructed by untrusted systems)
  are cast as intended, in particular, ballots encapsulate voters' choices. This 
  property is complimentary  to individual verifiability, which allows voters to 
  check whether their ballot is accepted by the bulletin board.}
hence, tampering with 
clocks might  be precluded by those assumptions.

OAuth tokens may contain timestamps~\cite[\S2.2]{RFC7662} and these can be used to determine 
the order in which ballots were authenticated. Similarly, ballots could be extended to 
include timestamps which can be used to determine the
order in which ballots were constructed.\footnote{%
  An encrypted choice comprises of El Gamal ciphertexts and non-interactive
  zero-knowledge proofs, and timestamps could be included in hashes used by proofs. This
  ensures that timestamps cannot be modified, if they are to be accepted by the bulletin
  board, because the board checks validity of proofs before accepting them.
}
These timestamps can be used by the bulletin board to patch the vulnerability. Indeed, rather than
archiving any encrypted choice previously accepted for a voter, the bulletin board can 
archive any encrypted choice associated with an earlier timestamp. 

The validity of tokens can only be checked by the bulletin board, because
tokens must remain secret. Thus, the bulletin board might convince itself that ballots
are authenticated, but it cannot convince other parties. Developing an authentication 
mechanism that permits anyone to check the validity of authentication tokens, rather than 
just the bulletin board, would be an interesting direction for future work. 
Alternatively, 
%ballots could be authenticated in a verifiable manner using cryptography.
voters can be issued with credentials and cryptography can be used to ensure that 
only voters can construct authorised ballots (i.e., authorised ballots are 
unforgeable~\cite[\S1]{Smyth18:voting-tutorial}).
For instance, Quaglia \& Smyth~\cite{2018-voting-authentication} replace 
tokens with digital signatures. But, solutions 
reliant on cryptography seem to require expensive infrastructures for voter 
credentials and %\sout{rarely address}
seemingly ignore the problem of corruption during the registration 
procedure~\cite[\S4]{2014-election-verifiability}. 
%This implicitly introduces an assumption that a \third\ party manages registration, 
%hence, solutions using cryptography do not necessarily reduce trust assumptions.
%Eliminating such trust assumptions is an interesting direction for future work.
Indeed, auditing is required to check whether any non-voters are issued credentials.
Eliminating such audits is desirable, but perhaps impossible.

\section{Related work}

Smyth \& Pironti%~\cite{2013-truncation-attacks-to-violate-beliefs,2014-truncation-attacks-to-violate-beliefs} 
~\cite{2013-truncation-attacks-to-violate-beliefs}
identify a flaw in Helios's sign-out procedure which can be exploited by TLS truncation attacks to
dupe voters into believing they have successfully signed-out, when they have not. Thus, an 
adversary can make a choice on the voter's behalf from the terminal used by the voter, thereby 
violating eligibility. Beyond eligibility and non-reusability, malleability has been exploited to launch violate
ballot secrecy~\cite{Smyth11:Helios,Smyth11:ReplayAttacks,Smyth12:Replay,Smyth12:Helios,Smyth15:ballotSecrecy} and verifiability~\cite{2014-election-verifiability}, 
and unsound proofs of knowledge have been exploited to violate verifiability~\cite{Bernhard12:Helios}.

\begin{comment}
Helios is reliant on a \trusted\ party to assure eligibility and non-reusability.
Although that party might convince itself that these properties are upheld, 
the ability to transfer that assurance is limited to the party's ``good word 
or sworn testimony"~\cite{Neff03}. Alternatively, these properties can be assured
cryptographically. E.g., Quaglia \& Smyth~\cite{Smyth18:HeliosC} define a transformation
from voting systems reliant on a \trusted\ party to voting systems reliant on cryptography,
moreover, they apply their transformation to Helios. But, solutions reliant on cryptography
seem to require expensive infrastructures for voter credentials and rarely address the
problem of corruption during the registration procedure, which implicitly introduces
an assumption that a \trusted\ party manages registration. Hence, solutions using 
cryptography to enable verifiable ballot authentication do not necessarily 
offer an advantage over schemes reliant on non-verifiable ballot authentication 
mechanisms, because both ultimately rely on a \sout{trusted} third party. Eliminating 
such trust is a direction for future work.
\end{comment}

\section{Conclusion}

We have shown that Helios does not satisfy strong non-reusability, because 
an adversary can cause a ballot other than a voter's last 
to be tallied. In particular, the adversary can intercept 
the authorisation token associated with the ballot that the
adversary wants tallied, wait until the voter has casts
their last ballot, and then release the intercepted token. 
The released token causes the bulletin board to accept the 
ballot that the adversary wants tallied, and to archive the 
voter's last ballot. Thus, strong non-reusability is not satisfied.
We have also shown that an adversary can %unduly influence election outcomes. 
choose the contents of such ballots.
In particular,
the adversary can exploit the educational needs of voters
to cast a ballot for the adversary's choice, and cause that ballot
to be tallied rather than the voter's last, as we have explained.
Consequently, adversaries can unduly influence
the selection of representatives. 
Although victims may detect malice, there is no evidence that malpractice
has taken place, hence, victims have little recourse.
The vulnerability is due to the manner in which Helios interacts
with  OAuth. 
Hence, our exploit should generalise to other systems that use OAuth
in a similar manner and to systems that use similar authentication 
mechanisms. We hope that this article leads to improvements in the
Helios election system, advances understanding of authentication
mechanisms, and helps system developers to integrate authentication 
mechanisms securely.

\section*{Acknowledgements}

We thank Elizabeth Quaglia, Susan Thomson and our anonymous reviewers 
for feedback that helped improve this paper. 
Smyth's work was partly performed at INRIA, with support from the European Research Council 
under the European Union's Seventh Framework Programme (FP7/2007-2013) / ERC project \emph{CRYSP} 
(259639).

%\section*{References}
\bibliographystyle{elsarticle-num}
\bibliography{main-eligibility}

\end{document}